\input harvmac.tex

\def\s{\sigma}
\def\be{\beta}
\def\avd#1{\overline{#1}}
\def\avdint#1{\overline{#1}^{\, '}}
\def\avdext#1{\overline{#1}^{\, ''}}
\def\avdjh#1{{\overline{#1}^{J,h^{(0)} }}}
\def\avdhI#1{{\overline{#1}^{h^{(1)} }}}
\def\avdhkI#1{{\overline{#1}^{h^{(k-1)} }}}
\def\avdhk#1{{\overline{#1}^{h^{(k)} }}}
\def\avdhs#1{{\overline{#1}^{h^{(s)} }}}
\def\tf{\tilde{f}}
\def\tk{\tilde{k}}

\def\th{\tilde{H}^{(n)}}
\def\thc{{\Omega_{clust}}^{(n)}}

\def\ch{{H}^{(n)}}

\def\avt#1{\langle#1\rangle}

\nfig\clust{Decomposition of a two-dimensional spin lattice in clusters
of $n_\s =12$ spins, with $n_b = 8$ boundary spins (white circles),
4 internal spins (black circles) and $n_J=16$ internal bonds (full lines)
per each cluster. The dashed lines represent the bonds between spins
belonging to different clusters.}

\nfig\enlib{Free energy as function of the temperature $T$ 
 in $d=2$ dimensions:
replica symmetry and one breaking solutions for the four spins
plaquette (dashed lines),
$SK$ replica symmetry and $SK$ one breaking (dot-dashed lines),
single plaquette (full line) with no boundary fields ($q^{a b}=0$).
The vertical bars represent the numerical error on the one breaking solution
for the plaquette.} 

\nfig\param{Order parameters for the
four spins plaquette in $d=2$ dimensions,
as function of the reduced temperature $T/T_{cr}$ :
a) replica symmetry $q_0$ and one breaking $q_1$; 
b) one breaking $m$.
}

\lref\hsI{N. Hatano and M. Suzuzi, J. Stat. Phys. {\bf 63}, 25 (1991)}
\lref\hsII{N. Hatano and M. Suzuzi, J. Stat. Phys. {\bf 66}, 897 (1992)}
\lref\sk{D. Sherrington and S. Kirkpatrick, Phys. Rev. Lett. {\bf 35}, 1792 (1975)}
\lref\book{M. Mezard, G. Parisi and M. Virasoro, 
{\it Spin Glass Theory and Beyond}\hfill\break (World-Scientific, Singapore, 1988)}
\lref\kar{M. Kardar and A. N. Berker, Phys. Rev. B {\bf 26}, 219 (1982)}
\lref\fish{M. E. Fisher, Phys. Rev. {\bf 176}, 257 (1968)}
\lref\gib{M. Serva and G. Paladin, Phys. Rev. Lett. {\bf 70 }, 105 (1993)}
\lref\bpI{G. Paladin and M. Serva, J. Phys. A {\bf 29 }, 1381 (1996)}
\lref\bpII{M. Serva and G. Paladin, Phys. Rev. E {\bf 54 }, 1 (1996)}
\lref\lib{A. Crisanti, G. Paladin and A. Vulpiani, 
{\it Products of random matrices in statistical physics} \hfill\break
 (Series in Solid State Sciences 104,  Springer-Verlag) (1993)}
\lref\ose{V.I. Oseledec,  Trans. Moscow. Math. Soc. {\bf 19}, 197 (1968)}
\lref\mat{T.D. Schultz, D.C. Mattis and E.H. Lieb, Rev. Mod. 
Phys. {\bf 36}, 856 (1964)}
\lref\sha{R. Shankar and G. Murthy, Phys. Rev. B {\bf 36}, 536 (1987)}
\lref\ser{M. Serva, Exact solution of a 2d random Ising model (submitted)}
\lref\serv{M. Serva, 2$d$ Ising model with layers of quenched spins (submitted)}
\lref\ons{L. Onsager, Phys. Rev. {\bf 65}, 117 (1944)}
\lref\bet{H. Bethe, Proc. R. Soc. London {\bf 150}, 552 (1935)}
\lref\pei{R. Peierls, Proc. R. Soc. London {\bf 154}, 207 (1936)}
\lref\vill{J. Villain, J. Phys. C {\bf 10}, L537 (1977)}
\lref\parI{G. Parisi, Phys. Lett. A {\bf 73}, 203 (1979)}
\lref\parII{G. Parisi, J. Phys. A {\bf 13}, L115 (1980)}
\lref\parIII{G. Parisi, J. Phys. A {\bf 13}, 1101 (1980)}
\lref\mezI{M. M\'ezard, G. Parisi, N. Sourlas, G. Toulouse and M. A. Virasoro,
J. Phys. {\bf 45}, 843 (1984)}
\lref\mezII{M. M\'ezard and M. A. Virasoro, J. Phys. {\bf 46}, 1293 (1985)}
\lref\nosI{G. Paladin, M. Pasquini and M. Serva, J. Phys. France I {\bf 4},1597 (1994)}
\lref\nosII{M. Pasquini, G. Paladin and M. Serva, Phys. Rev. E {\bf 51}, 2006 (1995)}
\lref\nosIII{G. Paladin, M. Pasquini and M. Serva, Int. J. Mod. Phys. B {\bf 9}, 399 (1995)}
\lref\nosIV{G. Paladin, M. Pasquini and M. Serva, J. Phys. France I {\bf 5}, 337 (1995)}
\lref\nosV{S. Scarlatti, M. Serva and M. Pasquini, J. Stat. Phys. {\bf 80}, 337 (1995)}
\lref\tou{G. Toulouse, Commun. Phys. {\bf 2}, 115 (1977)}
\lref\zoo{S. Kobe and T. Klotz, Phys. Rev. E {\bf 52}, 5660 (1995)}
\lref\che{H.-F. Cheung and W. McMillan, J. Phys. C {\bf 16}, 7027 (1983)}
\lref\fis{M. E. Fisher and W. Selke, Phys. Rev. Lett. {\bf 44}, 1502 (1980)}
\lref\kou{F. Koukiou, Europhys. Lett. {\bf 7}, 297 (1992)}
\lref\sau{L. Saul and M. Kardar, Phys. Rev. E {\bf 48}, R3221 (1993)}
\lref\tho{M. F. Thorpe and D. Beeman, Phys. Rev. B, {\bf 14}, 188 (1976)} 
\lref\van{J. L. Van Hemmen and R. G. Palmer, J. Phys. A, {\bf 15}, 3881 (1982)}

\vglue 2 truecm

\centerline{\bf A VARIATIONAL APPROACH } 
\centerline{\bf TO ISING SPIN GLASSES IN FINITE DIMENSIONS}
\vskip 1.4truecm\noindent 
\centerline{R. Baviera$^{(1,3)}$, M. Pasquini$^{(2)}$ and M. Serva$^{(2,3)}$} 
\vskip .5truecm \centerline{$^{(1)}$\it Dipartimento di Fisica, Universit\`a dell'Aquila}
\centerline{$^{(2)}$\it Dipartimento di Matematica, Universit\`a dell'Aquila}
\centerline{$^{(3)}$\it Istituto Nazionale Fisica della Materia, Universit\`a dell'Aquila}
\centerline{\it I-67010 Coppito, L'Aquila, Italy}
\vskip 1.6truecm
\centerline{ABSTRACT}

We introduce a hierarchical class of approximations of the
random Ising spin glass in $d$ dimensions.
The attention is focused on finite clusters of spins where
the action of the rest of the system is properly taken into account.
At the lower level (cluster of a single spin) our approximation
coincides with the $SK$ model while at the highest level it 
coincides with the true $d$-dimensional system.
The method is variational and it uses the replica approach to
spin glasses and the Parisi ansatz for the order parameter. 
As a result we have rigorous bounds for the quenched free energy 
which become more and more precise when larger 
and larger clusters are considered.

\vskip .4truecm

\hfill\break
\noindent
PACS NUMBERS: 05.50.+q, 02.50.+s 

\vfill\eject

\newsec{Introduction}

Research around spin glasses in finite dimensions is very active 
since it is still unclear if they share all the
qualitative features of the mean field model ($SK$).
Since a direct study of these systems is quite complicated both
from a numerical and an analytic point 
it could be of some interest to consider
corrections to the $SK$ model which 
partially take into account of the dimensionality.
Our aim is to find out a systematic and rigorous way to introduce 
these corrections.
As a result, we generate a class of models which interpolate between the
mean field $SK$ model and exact spin glasses in finite dimensions.
Our approach uses the replica formalism together with the 
celebrated Parisi ansatz for the order parameter.

The standard replica approach to the $SK$ model reduces
the problem to a single spin whose replicas interact 
via the variational order parameters $\{ q^{a b} \}$ that can be 
thought as 'coupling fields'.
This is the analogue of what one has 
for the mean field model of the ordinary ferromagnetic
Ising systems. In this second case, in fact, 
one has a single spin in a magnetic field generated by 
the rest of the system.

Both models, $SK$ and mean field Ising model, 
can be regarded as an approximation
of the associated Ising system in finite $d$ dimensions, but in both cases
any reference to the dimensionality is lost.
The approximation can be improved and a memory of the dimensionality
can be maintained if, in spite of considering a single spin in a bath, 
one focus the attention on a cluster of interacting spins
in a bath generated by the rest of the system. 
The strategy, which is very successfully applied for ordinary spin systems
(Bethe-Peierls approximation \bet -\pei ),
has been recently extended to spin glasses \bpI -\bpII . 
Actually the approach of \bpI -\bpII\
turns out to be not too much effective, since it 
does not allows for a study of the replica symmetry breaking.
This fact reduces the scope of the method
to low dimensional spin glasses, while for $d \ge 3$ dimensions 
it fails in describing the most striking feature of these systems.

In this paper we introduce a new approach which 
allows for symmetry breaking.
The attention is focused on finite clusters of spins where
the action of the rest of the system is properly taken into account.
The approximations we obtain are organized
hierarchically according to the size of the clusters.
At the lower level (cluster of a single spin) our approximation
coincides with the $SK$ model while at the highest level it 
coincides with the true $d$-dimensional system.
The method is variational and it uses the replica approach to
spin glasses and the Parisi ansatz for the order parameter. 
As a result we have rigorous bounds for the quenched free energy 
which become more and more precise when larger 
and larger clusters are considered.

Let us briefly sum up the contents of the paper.

In section 2, we introduce the model and we generalize 
the standard replica approach in order to take advantage
from the cluster partition of the lattice.

In section 3, we derive the new variational approach 
and we find out analytic lower bounds of the free energy 
of the $d$-dimensional spin glass.

In section 4, we choose the Parisi ansatz in order to obtain
a computable solution of the problem. In particular, 
we write down the free energy in the case of $k$ 
symmetry breaking.

In section 5,
we test our method against of the case of a 
plaquette of four spins in $d=2$ dimensions;
the free energy and the order parameter are obtained
at all the temperatures
for the replica symmetry and one symmetry breaking solutions.

In section 6,
we discuss some aspects of our approach which seems
to be useful for improving usual Monte Carlo simulations
for spin glasses in finite dimensions.

\newsec{New look at the replica approach}

We consider Ising spin glass models with nearest neighbours interactions
on a $d$-dimensional lattice of $N$ sites. The hamiltonian is
$$
H = - {1\over(2 d)^{1\over2}} \ \sum_{  (i,j) } J_{i,j} \s_i \s_j
$$
where the $\{ \s_i = \pm 1 \}$ are the $N$ spin variables
and the $\{ J_{i,j} \}$ are the $dN$
independent normal gaussian random variables
(zero mean and unitary variance). 
The sum runs on all the $dN$ nearest neighbours sites $(i,j)$.

The partition function reads
$$
Z = \sum_{ \{ \s \} } \exp \{ - \be H \}
$$
where $\be$ is the inverse temperature. The quenched free energy is
\eqn\free{
f_d = - \lim_{N \to \infty} {1\over{\be N}} \avd {\ln Z}
}
where $\avd{ \ \cdot \ }$ indicates the average 
over the disorder variables $\{ J_{i,j} \}$.
Indeed, almost all the disorder realizations 
have the same free energy in the thermodynamic limit
$N \to \infty$.

Unfortunately, it is not possible to find out an explicit expression of
\free\ in terms of simple functions
because of the presence of the logarithm in the
disorder average. The standard replica approach \book\ tries to avoid 
this difficulty replacing the above quenched average with
the annealed average of the $n$-th power of the partition function $Z$ with
integer $n$.
In fact, if the result can be analytically continued to real $n$, one has

\eqn\rep{
f_d = -  \lim_{n \to 0} \ \lim_{N \to \infty} \ {1\over{\be N n}} \ln \avd {Z^n}
}

The average over the gaussian disorder variables gives
\eqn\zn{
\avd{Z^n} = \exp \left\{ {1\over4} \be^2 N n \right\} 
\ \ \sum_{ \{ \s \} } \exp \left\{ 
{\be^2 \over 2d} \sum_{  (i,j) } \sum_{a<b} \s_i^a \s_i^b \s_j^a \s_j^b
\right\}
}
where $\s_i^a$ is the $a$-th replica of the spin in the $i$-th site.

Unfortunately, even in the replica context the free energy can be 
computed only in the infinite dimension limit.
In this case, in fact, one reduce to the celebrated $SK$ model \sk ,
and one has 
\eqn\fsk{
 f_\infty \  =  \
 - {\be\over4}   + \lim_{n \to 0} \ {1\over{n}} \ \max_{ \{ q^{a b} \} } \left[
\ {\be\over2}  \sum_{a<b} (q^{a b})^2
- {1\over\be} \ln \  \sum_{ \{ \s \} }
\exp \left\{ \be^2  \sum_{a<b} q^{a b} \s^a \s^b \right\} \ \right]
}
where $q^{a b}$ is a real matrix.
In the limit $n \to 0$ this maximum is found following the Parisi ansatz
\refs{\parI\parII{--}\parIII}.
When $d$ is finite, no analogous results are available, so that
it is sensible to look for approximations as the one
in this paper.

All the above expressions are so classical that it could appear completely
useless to have reproduced them here,
indeed, the reason is 
that we would like to recast them in a more general form
introducing the notion of cluster partition of the set of the $N$ spins.
The new formulation, which is more general and provides the 
technical ingredients for our variational approach, 
reduce to the standard replica trick in the case of clusters of a single spin.

To have an idea of the clusters we have in mind
think to a plaquette of four nearest-neighbours spins in two dimensions, 
or a cube of eight spins in three dimensions.
In general, we perform a decomposition of the set of the spins into 
clusters of the same shape, such that each spin belongs to one and 
only one of them.
In the following we indicate with $(i,j)^{'}$
all the couples of nearest-neighbours sites
that belong to the same cluster, and 
with $\avdint{ \ \cdot \ }$ 
the disorder average over the couplings
between them. In the same way $(i,j)^{''}$ denotes all the nearest-neighbours sites
of different clusters, and $\avdext{ \ \cdot \ }$ the related disorder average.
Finally, $(i)^{'}$ runs only over the boundary sites of all the clusters.
Moreover, the following definitions are useful
$$
\left\{ \eqalign{
& n_\s = {\rm number \ of \ spins \ in \ a \ cluster} \cr
& n_b = {\rm number \ of \ boundary \ spins \ in \ a \ cluster} \cr
& n_J = {\rm number \ of \ bonds \ in \ a \ cluster} \cr
}\right.
$$
which imply that $N\over n_\s$ is the 
total number of clusters in the system, and that  
$$
\left\{ \eqalign{
& \sum_{  (i)^{'} } 1 = {n_b\over n_\s} N \cr
& \sum_{  (i,j)^{'} } 1 = {n_J\over n_\s} N \cr
& \sum_{  (i,j)^{''} } 1 = \left( d - {n_J\over n_\s} \right) N \cr
}\right.
$$

For instance,  in \clust\ one has clusters of $n_\s =12$ spins, with
$n_b = 8$ boundary spins and $n_J=16$ internal bonds per each cluster.

Than, we compute again the free energy with the replica trick,
but this time we perform the annealed
average only over those bounds that couple different clusters:
\eqn\freeclust{
f_d = - \lim_{N \to \infty} \ \lim_{n \to 0} \ {1\over{\be N n}} 
\ \avdint{\ \ln \ \avdext {\ Z^n \ }\ }
}
Somehow, this expression interpolate between \rep\ which corresponds to
clusters of a single spin (no couplings inside the clusters)
and the quenched expression \free\
which correspond to a single cluster of size of order $N$ spins.

An easy calculation gives
\eqn\znII{
\avdext{Z^n} = \exp \left\{ {1\over4} \be^2 N n \left( 1 -{n_J \over d \ n_\s} \right) \right\} 
\ \ \sum_{ \{ \s \} } \exp \{ - \be \ch \}
}
where
\eqn\hcap{
\ch = - {1\over(2 d)^{1\over2}} \sum_{  (i,j)^{'} } J_{i,j} \sum_{a=1}^n \s_i^a \s_j^a \ 
- {\be \over 2d} \sum_{  (i,j)^{''} } \sum_{a<b} \s_i^a \s_i^b \s_j^a \s_j^b
}
Remark that the first sum, which runs on internal couplings,
 disappears when the clusters are of a single spin.
In this case \znII\ and \hcap\ reduce to \zn .

\newsec{The variational approach}

Let us start by only considering clusters where all boundary spins are 
topologically equivalent, as for example a $d$-dimensional hyper-cube
of $2^d$ spins, or the crosses shown in \clust\ on a two-dimensional
lattice.

We now introduce a trial hamiltonian $\th$
instead of $H^{(n)}$, where the first term related to the interactions 
between spins of the same cluster is left unchanged, while the second is
modified with the replacement 
\eqn\sub{
\s_i^a \s_i^b \s_j^a \s_j^b \ \to \
q^{a b} \left( \s_i^a \s_i^b \ +\ \s_j^a \s_j^b \right)
}
where the $\{q^{a b} \}$ are a set of
variational parameters of the problem.
Let us recall that $i$ and $j$ are a couple of boundary sites
of different clusters.
The intuitive meaning of our approximation is clear:
the coupling field $\{q^{a b} \}$
simulates the action of the rest of the system over a the boundary
of a cluster
in the replica space.
Remark that now the spins on the
boundary of different cluster  do not interact, 
so that the total hamiltonian is the sum of 
the hamiltonians of each cluster.
Therefore, with the replacement \sub , the new hamiltonian $\th$ has the form
\eqn\Hnew{
\th = \sum_{clust} \thc
}
with
\eqn\Hclust{
\thc = - {1\over(2 d)^{1\over2}} \sum_{  (i,j)^{'} }^{clust} J_{i,j} 
\sum_{a=1}^n \s_i^a \s_j^a \ 
- {\be \over n_b} \left( n_\s -{n_J \over d} \right)
\sum_{  (i)^{'} }^{clust} \sum_{a<b} q^{a b} \s_i^a \s_i^b
}
where now the sums $\sum_{  (i,j)^{'} }^{clust}$
 and $\sum_{  (i)^{'} }^{clust}$ run over, respectively, the internal
nearest neighbours bonds and the boundary sites of a single cluster.

Using the convexity of the exponential, the following inequality holds
for any integer $n>1$:
\eqn\dis{\eqalign{
\avdint{\ \ln \sum_{ \{ \s \} } \exp \{ - \be \ch \}} \ = & \
 \avdint{\ \ln \avt{ {\rm e}^ {- \be (\ch - \th) } }} \ +
 \  \avdint{\ \ln \sum_{ \{ \s \} }  \exp\{ - \be \th \}} \ \ge \cr
\ge  \ & \max_{ \{ q^{a b} \} } 
     \left[ - \be \avdint{\avt{ \ch - \th }} \ + \
 \avdint{\ \ln \sum_{ \{ \s \} } \exp \{- \be \th \} } \right]
}}
where the $\avt \cdot$ indicates the average 
over the Gibbs measure induced by the 
hamiltonian $\th$. Since the sites $i$ and $j$ 
belong to different clusters, one has
\eqn\overlap{
\avdint{\avt{\s_i^a \s_i^b \s_j^a \s_j^b}} = \
\avdint{\avt{\s_i^a \s_i^b}} \ \ \avdint{ \avt{\s_j^a \s_j^b}} = \
{\avdint{\avt{\s^a \s^b}}  }^2
}
where the indices have been suppressed because of the equivalence
of boundary spins.
As a consequence one can write the simple expression
$$
- \be \avdint{\avt{ \ch - \th }} = 
  {1\over2} \be^2 N \left( 1 -{n_J \over d \ n_\s} \right) 
   \sum_{a<b}  \left(  {\avdint{\avt{\s^a \s^b}}}^2
   - 2 \ q^{a b} \ \avdint{\avt{\s^a \s^b}}
\right)
$$

The maximum in the right hand side of \dis\ can be found
deriving it respect to each $q^{a b}$, so that after some trivial algebra
one has the following system of ${1\over2} n(n-1)$ self-consistent equations
\eqn\qab{
q^{a b} = \avdint{\avt{\s^a \s^b}}
\qquad \qquad 1 \le a  < b \le n}

The right hand side of \dis\ is the maximum of
an expression containing averages with respect to the Gibbs 
measure which are quite complicated. 
Fortunately, a more simple and compact expression
exists which has the {\bf same} maximum in the {\bf same} point
corresponding to the solution of \qab .
This expression is
\eqn\sost{
- {1\over2} \be^2  N \left( 1 -{n_J \over d \
n_\s} \right) \sum_{a<b} (q^{a b})^2  \ +
\  \avdint{\ \ln \sum_{ \{ \s \} }  \exp\{ - \be \th \}}  
}
Therefore, \sost\ can be used to replace the one
into the square parenthesis in the
right hand side of \dis .
Then, taking in mind that $\th$ is 
an hamiltonian fully decomposed into the
hamiltonians $\thc$ corresponding to the  $N\over n_\s$ different clusters, 
it is possible to perform the thermodynamic limit
and then the limit $n \to 0$.
In doing this second limit one has to be careful since the inequality
\dis\ has been established for integer $n>1$
and it changes direction 
when we perform the analytic continuation to
real $n<1$.
In conclusion, one has
\eqn\fdis{
f_d \ \ge \ \tf_d
}
with
\eqn\ffinal{
\eqalign{ \tf_d \ & \equiv  \ 
 - {\be\over4}  \left( 1 -{n_J \over d \ n_\s} \right) + \cr
 & +  \lim_{n \to 0}  \ {1\over{n}} \ \max_{ \{ q^{a b} \} } \left[ 
{\be\over2} \left( 1 -{n_J \over d \ n_\s} \right) \sum_{a<b} (q^{a b})^2
- {1\over\be n_\s} \ 
\avdint{\ \ln \sum_{ \{ \s \} }  \exp\{ - \be \Omega^{(n)}  \}}
\ \right] \cr }}
where $\Omega^{(n)}$ is a representative hamiltonian of a single cluster.

Before ending this section we would like to stress that
\ffinal , derived for clusters where the boundary spins are topologically
equivalent, can be easily extended to a generic cluster decomposition
of  the lattice (see Appendix).
In this case to every boundary spin $\s_i$ is associated a different
$q_i^{a b}$ and the maximization can become very complicated.
 Nevertheless, 
\fdis\ and \ffinal\ with a single $q^{a b}$
still hold although $\tf_d$ is not anymore the optimal
approximation.
The maximum is reached when
$$
q^{a b} = {1\over n_b} \sum_{\{ i \}^{'}} \avdint{\avt{\s^a_i \s^b_i}}
$$

Let us briefly sum up the results of this section.
We have found lower limits $\tf_d$ for the quenched free
energy $f_d$ of a spin glass in $d$ dimensions via the replica formalism.
The free energies $\tf_d$ approximate better and better the $f_d$
when the size of the cluster increases.
The structure of the solution is familiar, since we have to compute a 
maximum of
a function which depends
on a set of ${1\over2} n(n-1)$ variational parameters 
in the limit $n\to 0$.

Notice that
$\tf_d$ turns out to be 
a generalization
of the  expression \fsk\ for the $SK$ model free energy $f_\infty$.
In fact, independently of the dimension $d$,
$\tf_d$ reduces to \fsk\ when one chooses
a cluster of a single spin. The proof is trivial
since in this case one has $n_\s=1$, $n_b=1$ and $n_J=0$
so that
the first term in the hamiltonian $\th$ vanishes.
This fact is quite interesting since it 
implies that the well-known expression \fsk\
for the $SK$ model free energy represents in our scheme, so to speak, 
the zero-order approximation 
of the random  Ising spin glass in finite dimensions. 

It also should be remarked that
in the limit $d\to\infty$, independently on the size 
of the clusters, one reduces to the $SK$  model.

\newsec{Replica symmetry breaking with the Parisi ansatz}
It is quite simple to show that in the $SK$ model,
for any integer $n > 1$, 
the maximum in \fsk\ is reached  when all the $q^{a b}$ assume the same value
 \eqn\qrs{
q^{a b} = q_0 \qquad \qquad 1 \le a < b \le n
}
with $q_0\ge0$. 
This is the replica symmetry solution, but unfortunately it turns out to be
unstable and unphysical in the limit $n\to 0$
(for example, it has a 
negative zero temperature entropy).

Parisi has proposed a simple way \refs{\parI\parII{--}\parIII} 
to broke the above symmetry
between the $n$ replicas.
His choice is at the first stage to organize them
in $n\over{m_1}$ groups of $m_1$
replicas, and  to assume a $q^{a b}$ with two different values.
The larger value corresponds to
$a$ and $b$ belonging to the same group,
and the smaller one to $a$ and $b$ in different groups.
This strategy can be iterated repeating the same procedure for each
group and all its subgroups, so that 
the $k$-th order breaking can be written as
\eqn\parans{
q^{a b} = q_s
\qquad \quad  {\rm if} \ \left\{ \eqalign{  
\left[ a\over m_s \right] &= \left[ b\over m_s \right] \cr
\left[ a\over m_{s+1} \right] &\neq \left[ b\over m_{s+1} \right] 
} \right.
\qquad \quad  {\rm with} \ \left\{ \eqalign{  
1 \le a &< b \le n \cr
0 \le & s \le k
} \right.
}
where $[ \cdot ]$ means integer part. 
All the $\{ q_s - q_{s-1} \}$ are assumed to be non-negative
and it also assumed
$m_0\equiv n$ and $m_{k+1}\equiv 1$.

The above Parisi ansatz is straightforward
for integer $n$
if all the $\{m_s\}$ and the
$\{{m_s\over m_{s+1}}\}$ can be chosen as integers.
The intriguing point is that,  
after the analytic continuation to real $n$
in the limit $n\to 0$, the $\{q_s , m_s\}$ are treated 
as a set of $2k+1$ real variational parameters
with the constraint 
$$
0 \le\dots\le m_s\le m_{s+1}\le\dots \le m_{k+1}\equiv 1
$$
This constraint allows for a well-defined overlap probability.
We recall that it is sufficient to use few symmetry replica breaking 
(say $k=2$) to achieve
a solution of the $SK$ model with realistic behaviours (such as, 
$T=0$ free energy consistent with numerical simulations, 
or $T=0$ non-negative entropy).

The ansatz \parans\ can be easily adapted to 
our more general $\tf_d$. 
 The main difference with the $SK$ model 
is the presence of the coupling terms in the hamiltonian $\Omega^{(n)}$, 
but they do not mix different replicas,
so that the usual steps used for solving the $SK$ model
can be repeated.
Recalling the well-known trick of the Gaussian integral,
the solution of $\tf_d$ with $k\ge0$ breaking
can be written as
\eqn\fk{
\tf_{d,k} = \max_{\{q_s , m_s\}} \left[
 - {\be\over4} \left( 1 -{n_J \over d \ n_\s} \right) 
 \left( (1-q_k)^2 +  \sum_{s=1}^k m_s (q_{s-1}^2-q_s^2)\right)
+ f_k  \   \right]
}
with
\eqn\fkclust{
\hfill
f_k = - {1\over\be n_\s m_1} \ \avdjh{
\ \ \ln \ \ 
\avdhI{\left[ \ \dots \ \ 
 \avdhkI{\left[ \ \ \
\avdhk{\left[ \ Z_k  \ \right]^{m_k}}
\ \ \ \right]^{m_{k-1}\over m_k} \ }
\dots \ \right]^{m_1\over m_2} \ }
\ \ } , }
\eqn\zkclust{
Z_k \ = \ \sum_{\{\s\}} \exp \{ - \be H_k  \}
}
and
\eqn\Hkclust{
\eqalign{
H_k = - \ {1\over(2 d)^{1\over2}} 
 \sum_{  (i,j)^{'} }^{clust} & J_{i,j} \s_i \s_j \ + \cr
 - & \ {1\over n_b^{1\over2}} \left( n_\s -{n_J \over d}\right)^{1\over2} 
\sum_{(i)^{'}}^{clust} \s_i \left(
 q_0^{1\over2} h_i^{(0)}
+ \sum_{s=1}^k (q_s -q_{s-1})^{1\over2} h_i^{(s)} \right)
}}
Each of the $k+1$ averages $ \{ \avdhs{ \ \cdot \ } \}$ 
($0\le s \le k$)
contains $n_b$ independent normalized Gaussian fields $\{h_i^{(s)}\}$
acting only on the boundary spins of the cluster.
The set $\{h_i^{(0)}\}$ is the only one to appear in a 
quenched average $\avdjh{\ \cdot \ }$ 
together with the $n_j$ random couplings $\{J_{i,j}\}$ internal
of the cluster.
Notice that in the hamiltonian \Hkclust\ we
have replicated only the $n_\s$ spin variables of 
the cluster.

Equations
\fk\ - \Hkclust\ have the same structure of the Parisi solution of
the $SK$ model with $k$ replica symmetry breaking, except for 
a more general form of $H_k$. In particular, 
the Parisi solution for the $SK$ model can be recovered,
independently on the dimension $d$, 
choosing a cluster of a single spin.
For a larger cluster the Parisi solution only can be recovered
when $d \to \infty$.
In both cases, in fact, the first sum in \Hkclust\ disappears, and the
factor in front of the second sum equals one.

It is worth interesting that
the dependence of the solution from the
number of dimensions $d$ is purely algebraic,
once the shape of the cluster is fixed,
 so that the same algorithm
holds for every dimension $d$, which plays only the role of a parameter.

\newsec{An application in $d=2$ dimensions}
We check our method 
in $d=2$ dimensions choosing the elementary plaquette of four
nearest neighbours spins as the cluster, so that
$n_\s = n_b = n_J = 4$.
With this choice 
the replica symmetry solution (\fk - \Hkclust\ with $k=0$) reads:
\eqn\exrs{
\tf_0 =  \max_{q_0} \left[
 - {\be\over8} (1-q_0)^2  
- {1\over 4\be } \avdjh{\ \ln \sum_{\{\s\}} \exp \{ - \be
 H_0  \} } \   \right]
}
with
$$
H_0 = - {1\over2} \sum_{i=1}^4 J_i \s_i \s_{i+1} 
\  - {1\over 2^{1\over2}} 
q_0^{1\over2} \sum_{i=1}^4  h_i^{(0)} \s_i
$$
while the solution with one replica breaking ($k=1$) is:

\eqn\exone{
\tf_1 = \max_{\{q_0, q_1, m\}} \left[
 - {\be\over8} 
 \left( (1-q_1)^2 +  m (q_0^2-q_1^2) \right)
- {1\over 4 \be m} \avdjh{
\ln 
\avdhI{\left[  
\sum_{\{\s\}} \exp \{ - \be H_1  \}
 \right]^m } }
  \right]
}
with
$$
H_1 = - \ {1\over2} 
 \sum_{i=1}^4 J_i \s_i \s_{i+1} \ 
- \ {1\over 2^{1\over2}}  
\sum_{i=1}^4 \s_i \left(
 q_0^{1\over2} h_i^{(0)}
+ (q_1 -q_0)^{1\over2} h_i^{(1)} \right)
$$

It is obvious that \exone\ reduces to \exrs\ when $q_1=q_0$ and $m=0$.
The maximum in \exrs\ and \exone\ can be found out using
standard numerical methods.
For instance, deriving \exone\ with respect to $\{q_0, q_1, m\}$,
one can write down a set of self-consistent
equations which can be solved numerically.

The result is that the order parameters differ from 0
below a critical temperature $T_{cr} \sim 0.86$, that is sensibly
lower of the corresponding one of the $SK$ model ($T_{cr} = 1$).

In \enlib\ we plot the free energies $\tf_0$ and $\tf_1$
as a function of the temperature $T$ in the range $0 < T < T_{cr}$.
They are compared with the $SK$ results and with
the free energy of an isolated
plaquette with gaussian couplings and no boundary fields. 
Our free energies show a certain improvement respect 
to the $SK$ ones from a quantitatively point of view,
while the isolated plaquette badly describes the systems
below the temperature $T=0.7$.

In \param-a and \param-b are plotted, respectively, the $q_0$, $q_1$ and the $m$ 
order parameters of the one breaking solutions, as a function
of the reduced temperature $T/T_{cr}$.
The qualitative behaviors are very similar to the $SK$ corresponding
parameters.

\newsec{Conclusions}
Let us start this paragraph by  
a technical remark about the implementation 
of an algorithm able to find the maximum in \fk .
The expression \fkclust\ for $f_k$ suggests that the 
number of breaking $k$ is the main source for the algorithmic
complexity.
In fact, one has to compute first a quenched average over $n_J+n_b$
Gaussian variables (the $J$'s and the $h^{(0)}$'s); 
then an average over other $n_b$ variables (the $h^{(1)}$'s),
and so on. Using MonteCarlo algorithms this leads to a computing time $t$ for
$f_{clust}^{(k)}$ proportional to
$$
t \sim (n_J+n_b) \ n_b^k
$$
so that a unitary growth of the breaking number
corresponds to a big growth of $t$ 
which is amplified of a factor $n_b$.
On the contrary $t$ only has a polynomial dependence on 
$n_J$ and $n_b$,
so that it is less difficult to increase the size of the cluster.
Finally, the dimension $d$
is not significative,
since the complexity of the algorithm
does not depend on $d$.  

A second remark is that the hamiltonian \Hkclust\ and the free energy
\fk\ in case of replica symmetry correspond to a single replica
spin glass of finite size with gaussian magnetic fields at the 
boundary. The variance has to be chosen in order to feign at the
best the action of the rest of the system
(a similar approach has been proposed by Hatano and Suzuzi
(\hsI -\hsII ), where the variance is fixed by a self-consistent equation).

For these two reasons we believe that our approach 
could be used to 
improve the numerical simulations of spin glasses.
In fact, the numerical approach tries to understand the properties
of spin glasses in thermodynamical limit using finite size systems,
i.e. finite clusters with periodic boundary or open conditions.
In our replica symmetry context we save  this scheme
but we can take into account more carefully 
of the action of the rest of the system without increasing
too much the computing time. The ordinary numerical study
chooses zero variance of magnetic field at the boundary,
while we have a variance which can be optimized.
In conclusion, one should:

1) consider the finite size system and apply 
gaussian fields of variance $q_0$ at the boundary;

2) compute numerically the free energy and the overlap
for various values of $q_0$;

3) choose $q_0$ in order that it equals the overlap 
${1\over n_b} \sum_{\{ i \}^{'}} \avdint{\avt{\s^a_i \s^b_i}}$
(notice that $q_0 =0$ would correspond to the standard numerical study
with open boundaries).

Investigations about this numerical strategy are actually
in progress.

\bigskip
\noindent
{\bf Acknowledgements}
\medskip
\noindent
We acknowledge the financial support 
 ({\it Iniziativa Specifica} FI11) of the I.N.F.N., 
  National Laboratories  of Gran Sasso.
 We are very grateful to G. Parisi 
for useful discussions concerning the preliminary investigations
on the argument
and to F. Ricci 
for useful discussions
concerning numerical implementation of the
method.

\appendix{A}{}

In this appendix we derive a generalization of
formulae \fdis\ and \ffinal\ for the more general case of a
full decomposition of the lattice in equal clusters with 
no topological equivalence of boundary sites.

For instance, think to a square cluster of $L^2$ sites
in $d=2$ dimensions.
First of all,
the corner sites have two external bonds at variance with 
the unique external bond of the other boundary sites.
Furthermore, the external location of the site along the 
boundary also determine the strength of the interaction with external
spins. 
For these reasons, in general, the averaged overlap
$\avdint{\avt{\s_i^a \s_i^b}}$
depends on the boundary site $i$, and \overlap\ does not hold
anymore.
It follows that we have to modify the replacement rule \sub\
in order to take into account the topological differences between the
various boundary spins.

The boundary sites 
can be grouped in $n_b$ classes; each class consists of $N\over n_\s$
topologically equivalent sites, one per cluster.
We focus the attention on a given cluster (the reference cluster),
so that its $n_b$ boundary sites
${(k)}^{'}$ are to be the representative elements
of each class. Then, we introduce the function $k(i)$ which
associate the generic boundary site $i$ to its representative of the
reference cluster.
Two sites of the reference cluster, say $k(i)$ and $k(j)$, are 
'adjoint nearest neighbours' if the couple $i,j$ 
belongs to the set $(i,j)^{''}$, and
$(\tk\to k)$ indicates all the adjoint nearest neighbours $\tk$ of the site $k$.

At this point is straightforward to replace
$q^{a b}$ with 
a set of $n_b$ parameters $\{q_k^{a b} \}$, 
one per each representative $k$ site.
The total number of variational parameters is,
therefore, ${n_b\over2} n (n-1)$.

The replacement rule \sub\ 
for a couple of boundary sites can be now generalized as follows:
 \eqn\asub{
\s_i^a \s_i^b \s_j^a \s_j^b \ \to \
q_{k(j)}^{a b} \s_i^a \s_i^b \ +\ q_{k(i)}^{a b} \s_j^a \s_j^b
}
In other words, each replaced external interaction leaves 
a different memory.

With the replacement \asub\ the hamiltonian of a cluster reads
\eqn\ahclust{
\Omega^{(n)} = - {1\over(2 d)^{1\over2}} \sum_{  (i,j)^{'} }^{clust} J_{i,j} 
\sum_{a=1}^n \s_i^a \s_j^a \ 
- {\be \over 2 d} \sum_{a<b} \sum_{  (i)^{'} }^{clust} \s_i^a \s_i^b
\sum_{(\tk \to k(i) )} q_{\tk}^{a b}
}
and the averaged overlaps of two boundary sites of different clusters are
$$
\avdint{\avt{\s_i^a \s_i^b \s_j^a \s_j^b}} = \
\avdint{\avt{\s_i^a \s_i^b}} \ \ \avdint{ \avt{\s_j^a \s_j^b}} = \
\avdint{\avt{\s_{k(i)}^a \s_{k(i)}^b}} \ \ \avdint{ \avt{\s_{k(j)}^a \s_{k(j)}^b}}
$$

Formula \dis\ still holds for any integer $n>1$, so that
\eqn\adis{
\avdint{\ \ln \sum_{ \{ \s \} } \exp \{ - \be \ch \}} \
\ge \  \max_{ \{ q_k^{a b} \} } 
     \left[ - \be \avdint{\avt{ \ch - \th }} \ + \
 \avdint{\ \ln \sum_{ \{ \s \} } \exp \{- \be \th \} } \right]
}
where
$$
- \be \avdint{\avt{ \ch - \th }} = 
{\be^2 \over 4d} {N\over  n_\s}
\sum_{a<b}  \sum_{(k)^{'}} \avdint{\avt{\s_k^a \s_k^b}}
\sum_{(\tk \to k)} \left( \avdint{\avt{\s_{\tk}^a \s_{\tk}^b}}
 - 2 q_{\tk}^{a b} \right)
$$

The maximum in the right hand side of \adis\ can be found
by solving the following system of ${n_b\over2} n(n-1)$ 
self-consistent equations
\eqn\aqab{
\sum_{(\tk \to k)} q_{\tk}^{a b} = 
\sum_{(\tk \to k)} \avdint{\avt{\s_{\tk}^a \s_{\tk}^b}}
\qquad \qquad \left\{\eqalign{
1 \le &k \le n_b \cr
1 \le a  &< b \le n }\right.}
and the expression to be maximized in \adis\ can be replaced by
the following expression which has the same maximum in the same point: 
$$
-{\be^2 \over 4d} {N\over  n_\s}
\sum_{a<b}  \sum_{(k)^{'}} q_k^{a b} \sum_{(\tk \to k)} q_{\tk}^{a b}
\ + \ \avdint{\ \ln \sum_{ \{ \s \} } \exp \{- \be \th \} } 
$$

Finally, the analytic continuation to real $n\to 0$ gives
\eqn\afdis{
f_d \ \ge \ \tf_d
}
with
\eqn\affinal{
\eqalign{ \tf_d \ & \equiv  \ 
 - {1\over4} \be \left( 1 -{n_J \over d \ n_\s} \right) + \cr
 & +  \lim_{n \to 0}  \ {1\over{n}} \ \max_{ \{ q_k^{a b} \} } \left[ 
{\be \over 4d  n_\s}
\sum_{a<b}  \sum_{(k)^{'}} q_k^{a b} \sum_{(\tk \to k)} q_{\tk}^{a b}
- {1\over \be n_\s} \avdint{\ \ln \sum_{ \{ \s \} } \exp \{- \be 
\Omega^{(n)} \} } 
\ \right] \cr }
}
Notice that the $ \{ q_k^{a b} \} $ are not all different
if there are symmetric sites in the clusters.
For example, the sites on the four corners of a square plaquette
in $d=2$ dimensions will share the same overlaps.

Finally, if we look for the maximum of \affinal\ with the constraint
$$
q_k^{a b} \ = \ q^{a b}   \qquad \qquad \forall \ k \ =\ 1, \dots , n_b
$$
we reduces to formula \ffinal .
Therefore, in this context, \ffinal\  is a worse approximation, except 
all the boundary sites of the cluster are topologically equivalent.

\listrefs
\listfigs

\end